\documentclass[fleqn,usenatbib]{mnras}

\usepackage{newtxtext,newtxmath}

\usepackage[T1]{fontenc}

\DeclareRobustCommand{\VAN}[3]{#2}
\let\VANthebibliography\thebibliography
\def\thebibliography{\DeclareRobustCommand{\VAN}[3]{##3}\VANthebibliography}

\usepackage{graphicx}	
\usepackage{amsmath}	
\usepackage{caption}
\usepackage{subcaption}

\defcitealias{Dovekie}{Dovekie}
\defcitealias{Grayling24}{G24}
\defcitealias{Thorp21}{T21}

\title[The Inverse Distance Ladder Problem]{On The Stability Of $H_0$ And The Inverse Distance Ladder}

\author[Popovic and Sullivan]{
B. Popovic$^1$\thanks{Email: B.A.Popovic@soton.ac.uk}, M. Sullivan$^1$
\\
$^1$ School of Physics and Astronomy, University of Southampton, Southampton, SO17 1BJ, UK \\
}

\date{Accepted XXX. Received YYY; in original form ZZZ}

\pubyear{2026}

\begin{document}
\label{firstpage}
\pagerange{\pageref{firstpage}--\pageref{lastpage}}
\maketitle

\begin{abstract}
The `Inverse Distance Ladder' uses relative-distance measurements with type Ia supernovae (SNe Ia), anchored to an absolute distance scale from Baryon Acoustic Oscillations (BAO) and the cosmic microwave background (CMB), to provide an alternative measurement technique for the local expansion rate $H_0$. Using SNe Ia from the Dark Energy Survey and BAO measurements from the Dark Energy Spectroscopic Instrument, we show that the inverse distance ladder is unable to explain the Hubble Tension, anchored as it is to the CMB and its value of $H_0 = 67.4 \pm 0.5$ kms$^{-1}$ Mpc$^{-1}$. To do so, we first show that the suite of systematics considered in cosmology analyses with SNe Ia only move the inferred $H_0$ by $<0.1$kms$^{-1}$ Mpc$^{-1}$, and second, we investigate the scale of redshift-dependent magnitude changes necessary to change the inferred inverse distance ladder $H_0$ from $67$ to the local network of distance measurements value of $73$, and the impact that this would have on other cosmological inferences with SNe Ia. We find that a change of $d\mu/dz = 0.2$ mag would be necessary to infer an $H_0$ in concordance with local distance measurements, and that this $d\mu/dz$ value would result in a Flat $\Lambda$CDM $\Omega_M = 0.23$, $10\sigma$ discrepant with other cosmological probes, {indicating that the precision of SNe Ia measurements preclude the necessary redshift evolution for an $H_0$ of 74 kms$^{-1}$ Mpc$^{-1}$}. Therefore, we conclude that current SN Ia cosmology leaves little freedom for the inverse distance ladder to yield $H_0$ values significantly different from $67$ kms$^{-1}$ Mpc$^{-1}$.
\end{abstract}




\section{Introduction}\label{sec:Intro}

The difference between early-time and late-time measurements of the Hubble constant $H_0$ remains one of the most significant tensions in modern cosmology \citep*[see review of][]{Verde24}.  Measurements using the local distance network, combining geometric anchors with a range of stellar distance indicators, yield $H_0 = 73.04 \pm 1.04$ km\,s$^{-1}$ Mpc$^{-1}$ \citep{H0DN25}. This is in a $\sim5\sigma$ tension with the inferred value of $67.4 \pm 0.5$ km\,s$^{-1}$ Mpc$^{-1}$ from anisotropies in the cosmic microwave background (CMB; \citealp{Planck20}).

Evidently, the late- and early-time measurements do not agree. Either the standard model of cosmology is incomplete, or unknown or poorly modelled systematic effects are biasing one, or both, of the probes. In this paper, we examine the `inverse distance ladder', which uses Baryonic Acoustic Oscillations (BAO) distance measurements, anchored to the CMB sound horizon at the time of photon--baryon decoupling after recombination to set an absolute distance scale. SNe Ia can then be used as standardisable candles to measure the shape of the expansion history down to $z=0$ \citep{Aubourg15,Cuesta15}.

Using this technique, \citet{Aubourg15} found $H_0 = 67.3 \pm 1.1$ km\,s$^{-1}$ Mpc$^{-1}$ using SNe Ia from \cite{Betoule14} and BAO from the Baryon Oscillation Spectroscopic Survey (BOSS, \citealp{Anderson14}). Subsequent studies with different samples have since found consistent $H_0 \sim 67$ km\,s\,$^{-1}$ Mpc$^{-1}$ values: $H_0 = 67.8 \pm 1.3$ km\,s$^{-1}$ Mpc$^{-1}$ \citep{Macaulay19} and $H_0 = 67.19 \pm 0.65$ km\,s$^{-1}$ Mpc$^{-1}$ \citep{Camilleri24}.

While the initial implementation of the inverse distance ladder provided an interesting and novel test of the consistency of SN Ia distances, in this paper we show that this form of the inverse distance ladder will be unable to resolve the modern Hubble Tension due to the anchoring of the sound horizon of the CMB and its attendant value of $H_0 = 67.4 \pm 0.5$ km\,s$^{-1}$ Mpc$^{-1}$. Any SN Ia systematic that could then provide an inverse distance ladder measurement \textit{other} than $H_0 \sim 67$ km\,s$^{-1}$ Mpc$^{-1}$ is already immediately ruled out by current SN Ia data. In Section \ref{sec:Method}, we provide the polynomial expansion technique employed in this paper, alongside the specific distance measurements and values for the Dark Energy Spectroscopic Explorer (DESI) BAO and Dark Energy Survey (DES) 5-year SN Ia (DES-SN5YR) measurements. In Section \ref{sec:shortcomings}, we lay out potential SN Ia systematics that affect the inverse distance ladder technique, and assess their potential impact. The results of our fits are given in Section \ref{sec:Results}, followed by discussion and conclusions in Section \ref{sec:Conclusions}.

\section{Methodology and Data}\label{sec:Method}

To calculate the cosmographic expansion of the universe down to $z=0$, we use a Taylor expansion of the scale factor $a$ following \citet{Visser04} and \citet{Zhang17} that makes minimal assumptions about the underlying cosmological model. This was first used on DES-SN5YR data by \cite{Camilleri24}, and is parametrized by
deceleration
\begin{equation}\label{eq:decel}
    q = -\frac{1}{H^2}\frac{1}{a}\frac{d^2a}{dt^2},
\end{equation}
jerk,
\begin{equation}\label{eq:jerk}
    j = \frac{1}{H^3}\frac{1}{a}\frac{d^3a}{dt^3}, 
\end{equation}
snap,
\begin{equation}\label{eq:snap}
    s =  \frac{1}{H^4}\frac{1}{a}\frac{d^4a}{dt^4}, 
\end{equation}
and lerk,
\begin{equation}\label{eq:lerk}
    l =  \frac{1}{H^5}\frac{1}{a}\frac{d^5a}{dt^5}. 
\end{equation}

The Hubble parameter can then be expressed to fifth order, as:
\begin{equation}\label{eq:HubbleExpansion}
    H(z) = H_0(1+\mathcal{H}_1 z + \mathcal{H}_2 z^2 + \mathcal{H}_3 z^3 + \mathcal{H}_4 z^4 )
\end{equation}
where $\mathcal{H}_i$ are defined as:
\begin{equation}
    \mathcal{H}_1 = 1+ q_0,
\end{equation}
\begin{equation}
    \mathcal{H}_2 = \frac{1}{2}(j_0 - q_0^2),
\end{equation}
\begin{equation}
    \mathcal{H}_3 = \frac{1}{6}(3q_0^2 + 3q_0^3 - 4q_0j_0 - 3j_0 - s_0),
\end{equation}
and 
\begin{align} 
\begin{split}
    \mathcal{H}_4 = \frac{1}{24}(-12q_0^2 - 24q_0^3 - 15q_0^4 + 32q_0j_0 + 25q_0^2j_0 \\ + 7q_0s_0 + 12j_0 -4j_0^2 + 8s_0 + l_0).
\end{split}
\end{align}
The subscript $_0$ on $q,j,l,s$ denotes the $z=0$ value for each parameter.

Assuming spatial flatness, we can similarly express the luminosity distance as 
\begin{equation}\label{eq:luminosityexpansion}
    D_L(z) = z + \mathcal{D}_1 z^2 + \mathcal{D}_2 z^3 + \mathcal{D}_3 z^4 + \mathcal{D}_4 z^5.
\end{equation}
where $\mathcal{D}_i$ are given by
\begin{equation}
    \mathcal{D}_1 = \frac{1}{2}(1-q_0),
\end{equation}
\begin{equation}
    \mathcal{D}_2 = -\frac{1}{6}(1-q_0 - 3q_0^2+j_0),
\end{equation}
\begin{equation}
    \mathcal{D}_3 = \frac{1}{24}(2-2q_0-15q_0^2+5j_0+10q_0j_0 + s_0),
\end{equation}
and
\begin{align} 
\begin{split}
    \mathcal{D}_4 = \frac{1}{120}(-6 +6q_0 + 81q_0^2 + 165q_0^3 + 105q_0^4 + 10j_0^2
    \\ -27j_0 - 110q_0j_0 - 105q_0^2j_0 - 15q_0s_0 - 11s_0 - l_0 ).
\end{split}
\end{align}

We provide a derivation of the $\mathcal{D}_L(z)$ term when the assumption of spatial flatness is relaxed in Appendix \ref{sec:Appendix}. We also test this lower-order expansion and fit for the spatial flatness term $k$ later in the paper as a further test of the robustness of inverse distance ladder measurements. 

\subsection{DESI BAO}

As a high-redshift anchor for distances, we use the BAO measurements from the second DESI data release \citep{DESI25}. The baryon decoupling in the early universe left an imprint on the matter distribution of the universe, detectable through the distribution of matter measured via galaxy distributions. BAO measurements are dubbed `standard rulers' \citep{BlakeGlazebrook03, SeoEisenstein03, Linder03, McDonaldEisenstein07, Alam17, DESI25}, and here we make use of these standard rulers to break the SN Ia degeneracy between $H_0$ and the SN Ia absolute magnitude, $M_0$. DESI DR2 provides two sets of distance measurements: separation vectors for pairs of galaxies oriented parallel to the line-of-sight $D_H(z)/r_d$ and separation vectors for pairs oriented perpendicularly $D_M(z)/r_d$. Together, these measurements are measured from for well over 6 million galaxies spread across $0.1 < z < 4.2$. An additional, volume-averaged quantity $D_V(z)/r_d$ is quoted for certain redshift bins with low statistics or signal-to-noise ratio.

In both cases, BAO measurements employ a physical scale set by the sound horizon $r_s$ at the epoch of decoupling ($z = z_{\star} \approx1600$, $r_d \equiv r_s (z_{\star})$). Following \cite{Camilleri24} and \cite{LemoLewis23}, we use a Gaussian prior of $r_d\sim \mathcal{N}(147.46, 0.28)$, which is independent of late-time cosmology and modelled alongside the late Integrated Sachs-Wolfe effect, optical reionisation depth, CMB lensing, and foregrounding. 

We calculate the necessary Hubble and transverse distances $D_H(z)$ and $D_M(z)$ as 
\begin{equation}\label{eq:hubbledist}
    D_H(z,\Theta) = \frac{c}{H(z, \Theta)},
\end{equation}
and
\begin{equation}
    D_M(z,\Theta) = \frac{D_L(z,\Theta)}{1+z},
\end{equation}
where $H(z, \Theta)$ and $D_L(z, \Theta)$ are given in Equations \ref{eq:HubbleExpansion} and \ref{eq:luminosityexpansion}, respectively. The volume-averaged distance measurement $D_V(z)$, the dilation scale, is given as 
\begin{equation}
    D_V(z,\Theta) = (z D^2_M(z,\Theta) \times D_H(z,\Theta) )^{1/3}.
\end{equation}

We use the DESI BAO measurements in our fits by finding the minimum of 
\begin{equation}
    \chi^2_{\rm BAO}(r_d, \Theta) = \vec{\Delta}^T \mathcal{C}^{-1}_{\rm BAO} \vec{\Delta},
\end{equation}
where $\Delta(r_d, \Theta)$ is the difference between real measurements with BAO and the predicted distances of the given cosmographic model. $\mathcal{C}_{\rm BAO}$ is the DESI-provided covariance matrix. 

\subsection{SN Ia data: DES-Dovekie}

Our sample of SNe Ia is the recently-released `DES-Dovekie' sample from \cite{Popovic26}, containing 1820 likely SNe Ia. DES-Dovekie contains calibration updates from \cite{Dovekie} that are used to update the original results from \cite{DES5YR} and data release from \cite{Sanchez24}. DES-Dovekie covers a redshift range of $0.10$ to $1.13$ with 194 complementary SNe Ia from low-redshift samples in the literature. 

The DES-Dovekie data release includes the fitted SN Ia light curve parameters, standardised brightnesses, bias corrections, redshifts, and distance moduli. Without an external source of absolute calibration, the DES-Dovekie sample, and indeed \textit{all} SN Ia samples, is unable to break the degeneracy between $H_0$ and the SN Ia absolute brightness $M_0$. Distance ladder techniques break this degeneracy with sources of external calibration.

We calculate the necessary data vector as 
\begin{equation}
    \mu_{\rm dat}(M_0) = \mu_{\rm dat} - M_0,
\end{equation}
where $\mu_{\rm dat}$ is the final cosmological distance of a given SN Ia and $M_0$ is the absolute magnitude shared between all SNe Ia, and our free parameter in this work. The distance modulus for our theory component is calculated as 
\begin{equation}
    \mu_{\rm theory}(z,\Theta) = 5 \log_{10}(D_L(z,\Theta)/1\rm Mpc) + 25.
\end{equation}
The difference between the real SN Ia distances and the theoretical distance for a given SN Ia is 
\begin{equation}
    D(M_0, \Theta) = \mu_{\rm dat}(M_0) - \mu_{\rm theory}(z, \Theta)
\end{equation}
and we minimize 
\begin{equation}
    \chi^2_{\rm SN} (M_0, \Theta) = \vec{D}^{T} \mathcal{C}_{\rm SN}^{-1} \vec{D},
\end{equation}
where $\mathcal{C}_{\rm SN}$ is the inverse covariance matrix as provided by \cite{Popovic26}. 

\subsubsection{Pantheon+ and Union SN Ia datasets}

As an additional test of the consistency of the inverse distance ladder, we additionally fit the Pantheon+ \citep{Brout22,Scolnic22} and Union SN Ia distances \citep{Union}; more information can be found in their respective data releases and cosmology results.

\subsection{The Inverse Distance Ladder}

To find our best-fitting cosmography, we combine the likelihoods as 
\begin{equation}\label{eq:totalchi}
    \chi^2(M_0,r_d,\Theta) = \chi^2_{\rm BAO}(r_d, \Theta) + \chi^2_{\rm SN}(M_0, \Theta),
\end{equation}
and fit the two simultaneously using the dynamic nested sampling package \texttt{Dynesty} \citep{Skilling04, Higson18, Speagle20, Koposov23}. We make use of the inbuilt stopping function that sets the stop criteria at $d\log z = 0.010$, and use 500 live points.

We fit two separate orders of polynomial for our fiducial results: $\Theta = \{ H_0, q_0, j_0, s_0\}$ and $\Theta = \{ H_0, q_0, j_0, s_0, l_0\}$, to determine which polynomial expansion to use for our systematic tests. \cite{Visser04} does not provide the $l_0$ term when dropping the assumption of spatial flatness; therefore, we only fit $\Theta = \{ H_0, q_0, j_0, s_0, k/a_0^2\}$ when considering the case that $k\neq0$.

\section{SN Ia Systematics and The Inverse Distance Ladder}\label{sec:shortcomings}

The inverse distance ladder approach consistently returns measurements of the Hubble constant consistent with that inferred from the CMB. We now investigate whether any known systematic effects could lead to an alternative $H_0$ value. First, we test whether the SN Ia systematics accounted for in measurements of dark energy cosmology are capable of shifting the central inferred $H_0$. Second, we compute the $d\mu/dz$ value necessary to change the SN Ia distance moduli as a function of redshift so as to shift the inverse distance ladder $H_0$ away from the CMB value. We investigate if the required systematic $d\mu/dz$ is supported by current data. 

\subsection{SN Ia Systematics}

DES-SN5YR, as part of DES-Dovekie\footnote{https://github.com/des-science/DES-SN5YR}, includes additional sets of distances and covariances associated with its systematics analysis (Tables 6 and 7 in \citealp{Popovic26}). We fit our cosmography with these distances to test the impact of known systematic effects on the inferred $H_0$ value. We take the SN distances for each systematic from DES-Dovekie and fit them with our 5th-order polynomial -- assuming a flat universe -- which is the best fit to our nominal data (Section \ref{sec:Results}).

\subsection{A SN Ia Systematic Capable of Resolving the Hubble Tension}

We next explore whether the Hubble tension may be explained by an unidentified systematic effect that causes a redshift evolution in the SN Ia data. To test the magnitude of the redshift evolution necessary to return $H_0 = 74$\,km\,s$^{-1}$ Mpc$^{-1}$ using the inverse distance ladder, we perform a test by separating the SN Ia data at $z\leq0.10$ and adding a magnitude shift to this low-redshift data as

\begin{equation}\label{eq:step}
\mu_{\rm dat}(z)= 
     \begin{cases}
       \mu_{\rm dat}(z)+x &\quad\text{if } z\leq0.10 \\
       \mu_{\rm dat}(z)+0 &\quad\text{otherwise,} \\ 
     \end{cases}
\end{equation}
where $x$ is the magnitude shift starting at $-0.05$\,mag and increasing to $-0.35$\,mag in steps of $0.05$\,mag. At each magnitude shift, we fit our 5th-order polynomial as normal, and record the $H_0$ and $M_0$ values. 

We also repeat this test for a more smoothly-varying redshift evolution:
\begin{equation}\label{eq:smooth}
    \mu_{\rm mod}(z) = \mu + p\times z,
\end{equation}
where $p$ follows the same pattern as our step size, starting at $0.05$ and increasing to $0.35$.

As a further test of the precision of SN Ia cosmology, we use the quick-fitting program \texttt{wfit} from \cite{Kessler09,Kessler19} to fit the $-0.05$ and $-0.30$ mag shift values to determine the dark energy equation-of-state $w$; this is done with SN-only in a flat $\Lambda$CDM cosmology, and SN Ia with a CMB prior in $w$CDM cosmology.

\section{Results}\label{sec:Results}

\subsection{Nominal Results}

Table \ref{tab:nominalresults} shows the 4th- and 5th-order polynomial results for the nominal DES-Dovekie data. Calculating the Akaike Information Criteria AIC$ =2K - 2\ln(\mathcal{L})$, where $K$ is the number of degrees of freedom of the model, we find only a $\Delta$AIC$\sim2$ between the two polynomial fits, with the 5th order polynomial being weakly preferred over the 4th order \citep{Trotta08}. The assumption of spatial flatness does not change the result: when we do not fix $k=0$ we find $H_0 = 66.59 \pm 0.69$ km\,s$^{-1}$ Mpc$^{-1}$, nearly identical to the 4th-order result. 

\begin{figure}
    \centering
    \includegraphics[width=8cm]{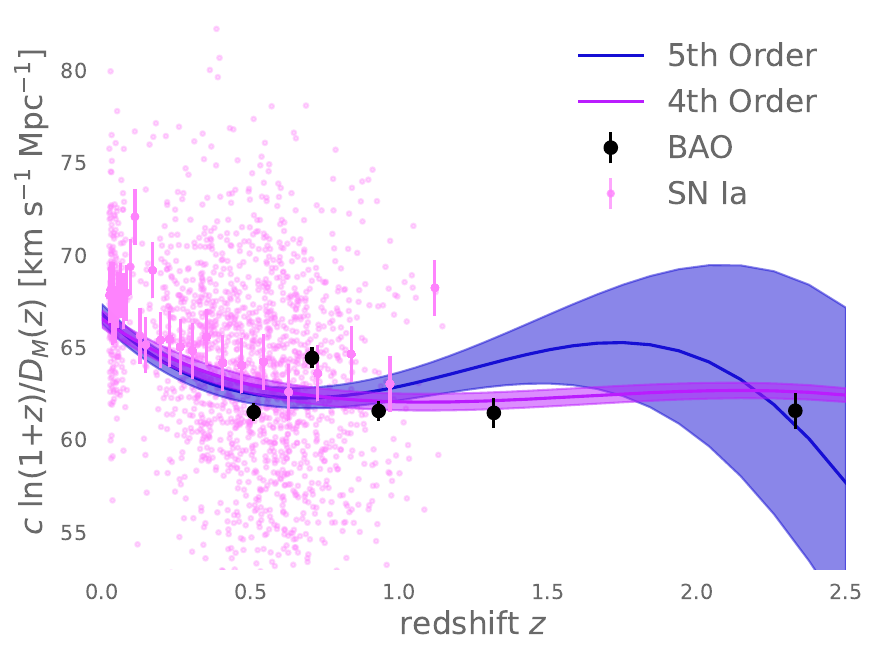}
    \caption{The $c\ln(1+z)/D_M(z)$ v redshift relationship, demonstrating the inverse distance ladder method. In black are the $D_M(z)/r_d$ measurements from DESI; individual SNe Ia are shown in light pink points and the binned averages in dark pink. Overplotted are the 4th (pink line) and 5th (blue line) order polynomial fits to the combined data. The y-intercept of this plot gives $H_0 \sim 67$, varying slightly with the choice of polynomial.}
    \label{fig:clnz}
\end{figure}

\begin{table}
    \caption{The 4th- and 5th-order polynomial results from the nominal DES-SN5YR (\lq DES-Dovekie\rq) data. We find a $\Delta$AIC of 2.1 between the two fits, indicating a weak preference for the 5th order fit. We also include the cosmographic expansion without the assumption of spatial flatness, denoted as $k \neq 0$.}
    \centering
    \begin{tabular}{c|ccc}
    Parameter & 4th Order & 5th Order & $k \neq 0$ \\
    \hline
    $H_0$ & $66.54 \pm 0.70$ & $66.82 \pm 0.74$ & $66.59 \pm0.69$ \\
    $M_0$ & $-19.35 \pm 0.02$ &  $-19.36 \pm 0.02$ & $-19.35\pm0.02$\\
    $q_0$ & $-0.35 \pm 0.01$ & $-0.48 \pm 0.03$ & $-0.35 \pm0.01$\\
    $j_0$ & $0.36 \pm 0.08$ & $0.89 \pm 0.08$ & $0.35 \pm 0.08$\\
    $s_0$ & $-0.54 \pm 0.05$ & $-0.17 \pm 0.27$ & $-0.54 \pm 0.05$\\
    $l_0$ & --  & $3.151 \pm 0.91$ & -- \\
    $k/a_0^2$   & -- & -- & $0.02\pm0.66$ \\
    \end{tabular}
    \label{tab:nominalresults}
\end{table}

Our nominal $H_0 = 66.82 \pm 0.74$ km\,s$^{-1}$ Mpc$^{-1}$ agrees within $1\sigma$ of the CMB value, and is consistent with previous inverse distance ladder measurements calibrated by BAO that are anchored to the CMB value \citep{Aubourg15,Macaulay19,Camilleri24}. Fig.~\ref{fig:clnz} shows the $c\ln(1+z)/D_M(z)$ value of our probes as a function of redshift; the y-intercept corresponds to $H_0$.

Additionally, we separate our BAO measurements into their line-of-sight component $D_H$ and perpendicular component $D_M$, and redo the 5th-order polynomial fit to test for consistency among the two distance measures. For the measurements that only contain $D_M$, we find 
$H_0 = 67.16^{+0.85}_{-0.80}$ km\,s$^{-1}$ Mpc$^{-1}$, compared to $H_0 = 71.2^{+2.0}_{-1.9}$  km\,s$^{-1}$ Mpc$^{-1}$ when considering only the $D_H$ BAO results. 

Our results are consistent when using the Pantheon+ and Union SN Ia data. For Pantheon+, we find $H_0 = 66.35^{+0.81}_{-0.86}$ km\,s$^{-1}$ Mpc$^{-1}$, and $H_0 = 66.15^{+0.86}_{-0.85}$ km\,s$^{-1}$ Mpc$^{-1}$ for Union.  

\subsection{The Shortcomings of the Inverse Distance Ladder as an Independent Determinant of $H_0$}

\begin{figure}
    \centering
    \includegraphics[width=9cm]{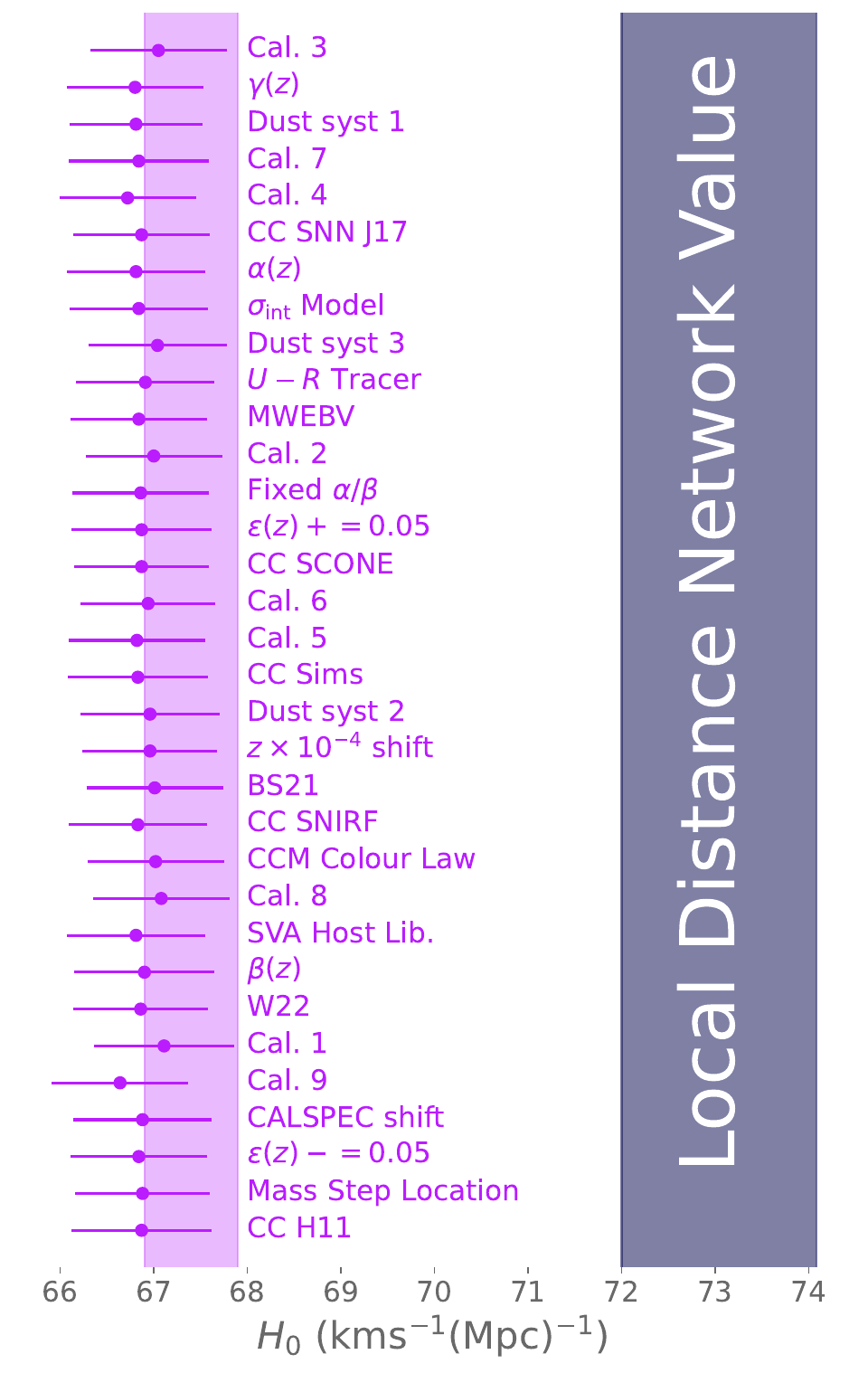}
    \caption{The SN Ia systematic effects tested in DES-Dovekie and their associated $H_0$ values from this work are shown in pink points with errors. The CMB value of $H_0 = 67.4 \pm 0.5$ is plotted in pink, and for reference the local distance network value of $H_0 = 73.04 \pm 1.04$  km\,s$^{-1}$ Mpc$^{-1}$ \citep{H0DN25} is plotted in grey. The scatter of $H_0$ values across all systematics is $0.1$ km\,s$^{-1}$Mpc$^{-1}$. For an explanation of the systematic names, see tables 6 and 7 in \citet{Popovic26}.}
    \label{fig:systematics}
\end{figure}

\begin{figure}
    \centering
    \includegraphics[width=9cm]{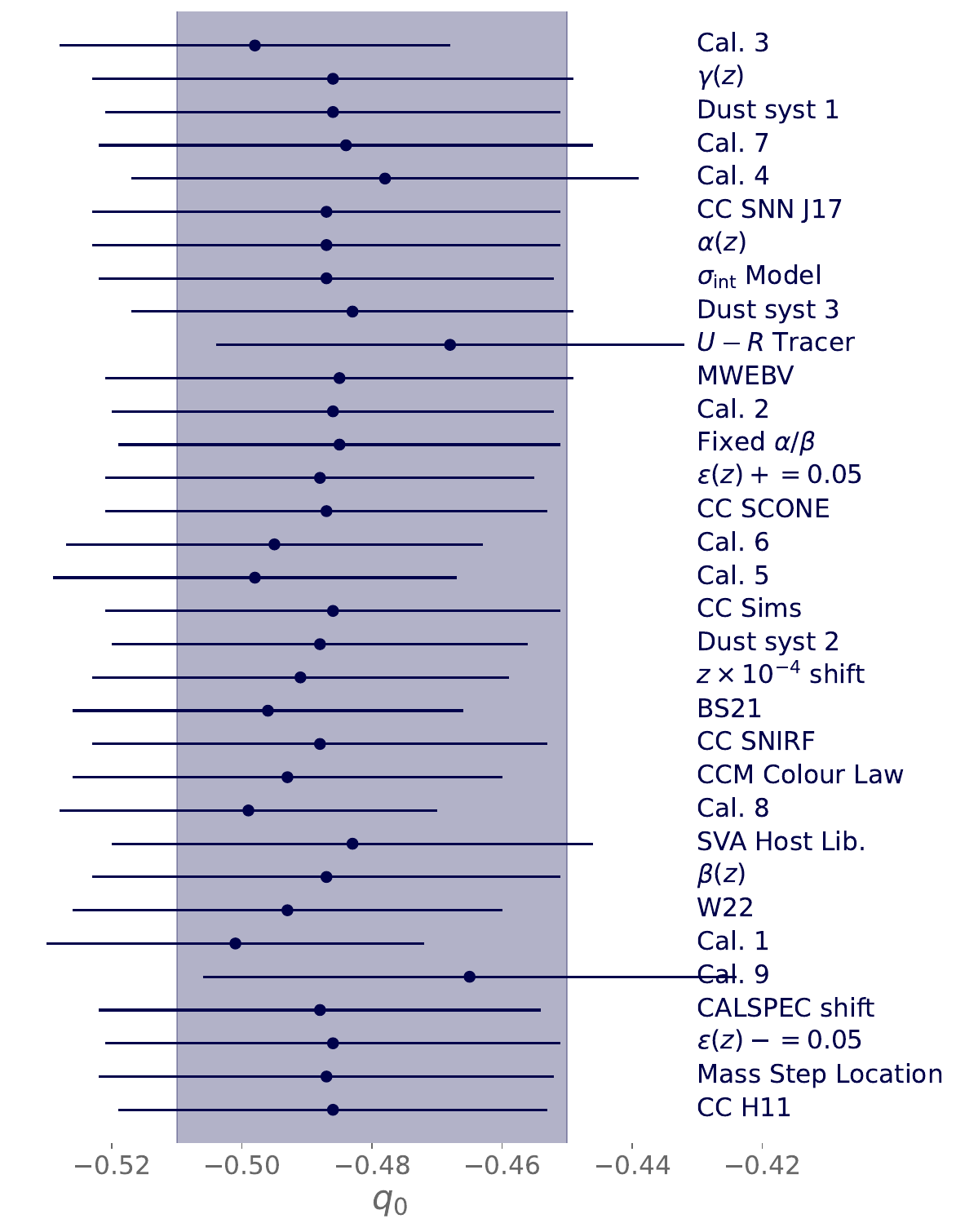}
    \caption{As in Fig.~\ref{fig:systematics}, but presenting the $q_0$ parameter. A $q_0 \sim 0.1$ corresponds to a universe without dark energy; $q_0 \sim -0.5$ signals a dark-energy dominant universe.}
    \label{fig:systematics_q0}
\end{figure}

Fig.~\ref{fig:systematics} shows the inferred $H_0$ value for each systematic tested in DES-Dovekie. We present the local distance network value \citep{H0DN25} in grey, and the CMB value in pink, for reference. The scatter across all SN Ia systematic effects is $0.1$km\,s$^{-1}$Mpc$^{-1}$: evidently, SN Ia systematic effects included in precision measurements of $w$ are not sufficient to resolve the Hubble Tension. Fig.~\ref{fig:clnz} shows the issue clearly: the absolute scale of the expansion history is set by the high-redshift BAO values, anchored to the CMB, and this value consistently results in inverse distance ladder measurements of $H_0 =\sim 67$  km\,s$^{-1}$ Mpc$^{-1}$. The $q_0$ parameter is equally robust to systematics of SNe Ia, varying from its nominal value of $-0.48$ by no more than 0.01, as in Fig.~\ref{fig:systematics_q0}. Dropping our assumption of flatness and allowing $k/a_0^2$ to float does not change our results; $H_0$ does not move from $66.5$ kms$^{-1}$ Mpc$^{-1}$ and the systematic spread remains less than $0.1$ km\,s$^{-1}$ Mpc$^{-1}$.

\begin{figure}
    \centering
    \includegraphics[width=8cm]{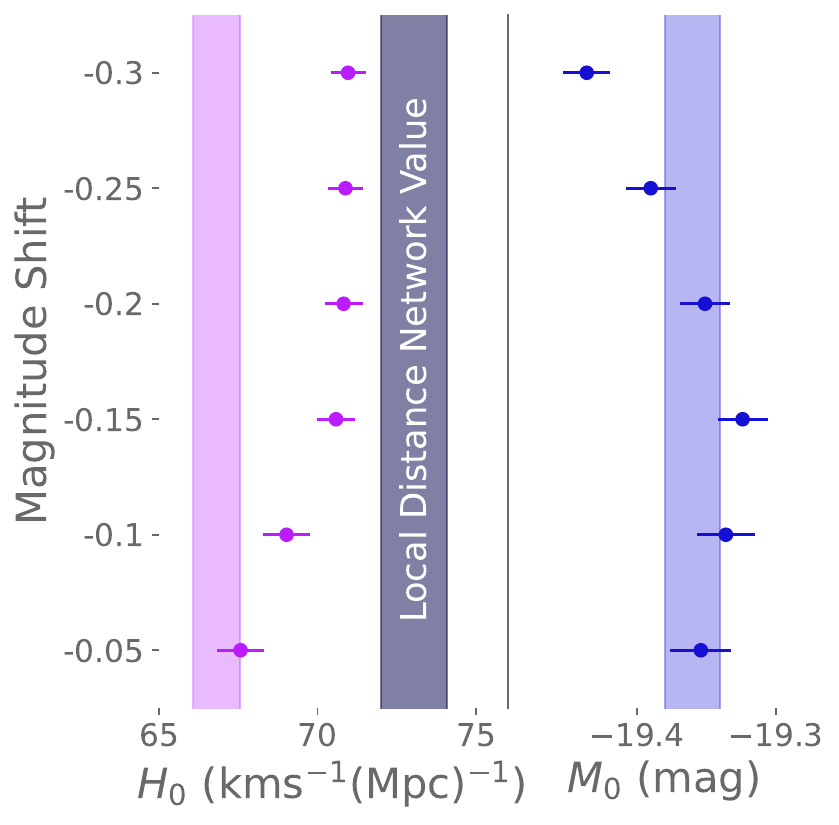}
    \caption{The inferred $H_0$ and $M_0$ values as a function of the change in SN Ia magnitudes below $z=0.1$. The $H_0$ values are shown in purple points, with the nominal value in pink fill, and the local distance network value \citep{H0DN25} provided in grey. Alongside we provide the equivalent $M_0$ values for the fit in blue points, and the nominal $M_0$ value in blue fill. }
    \label{fig:H0shifts}
\end{figure}

\begin{figure}
    \centering
    \includegraphics[width=9cm]{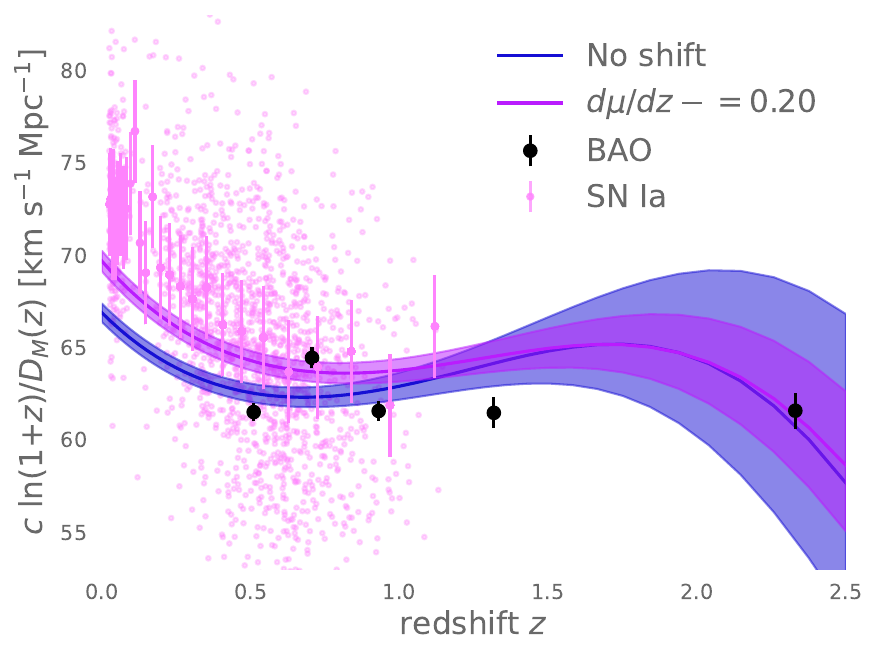}
    \caption{A $d\mu/dz = -0.20$ mag shift is applied to the SN Ia data in pink, such that $H_0 = 71$ km\,s$^{-1}$ Mpc$^{-1}$. The nominal cosmographic expansion is shown in blue, for contrast against the $d\mu/dz = -0.20$ mag slope in pink. A better cosmographic fit may be acquired with the addition of even higher-order terms, but would not change the resulting $w$CDM inference. }
    \label{fig:linear_change}
\end{figure}

Fig.~\ref{fig:H0shifts} shows the results of our systematic test of increasing the step applied to SN Ia magnitudes. A shift of $-0.3$ mag, a factor of $\times10$ greater than any SN Ia systematic considered in \cite{Popovic26}, only recovers $H_0 = 70.96^{+0.55}_{-0.54}$  km\,s$^{-1}$ Mpc$^{-1}$, still $\sim4\sigma$ away from the local distance network value of $73.4$  km\,s$^{-1}$ Mpc$^{-1}$. Fig.~\ref{fig:linear_change} shows the impact of the $p=-0.20$ slope, but such a slope is ruled out by other cosmological measurements. Even a small shift of $-0.05$ mag, which returns the same $H_0$, within error, of our nominal result, is disfavoured by the data with a $\Delta$AIC $=20$; this number only increases up to $\Delta$AIC $=4000$ for the $-0.30$\,mag shift value. The $M_0$ value, also fit, remains more stable around $-19.3$, only increasing outside of our nominal value for the $-0.30$ mag offset value. 

The situation is similar for the smooth $d\mu/dz$ shift in Equation\ref{eq:smooth}. A value of $p = 0.2$\,mag across the redshift range $0 < z < 1$ gives $H_0 = 71.08$  km\,s$^{-1}$ Mpc$^{-1}$ and a strongly-disfavoured fit; $p = 0.05$ does not change the $H_0$ value but does negatively impact the quality of the fit. 

Even though a magnitude shift of $x = -0.30$\,mag is not sufficient to recover $H_0 = 73$ kms$^{-1}$ Mpc$^{-1}$, it is still worth investigating other cosmological parameters. For a flat $\Lambda$CDM model, the nominal cosmology from DES-Dovekie is $\Omega_M = 0.330 \pm 0.015$; applying the $-0.05$\,mag shift for $z\leq0.10$ gives $\Omega_M = 0.28783$; $2.8\sigma$ from the nominal cosmology. For the shift of $x = -0.30$ mag, the difference is exacerbated, increasing to $13\sigma$ with $\Omega_M = 0.13398$. The $w$CDM values are similarly discrepant; the $x = -0.05$ mag shift gives $w= -1.053$ and $\Omega_M = 0.303$, close to our actual cosmology results, but the $x = -0.30$\,mag step gives $w = -1.499$ and $\Omega_M = 0.236$.

\section{Discussion and Conclusions}\label{sec:Conclusions}

Since its introduction in 2015, the inverse distance ladder has provided a consistent estimate of $H_0 \sim 67$ km\,s$^{-1}$ Mpc$^{-1}$, taking the absolute distance scale set by the CMB. Given the use of SNe Ia in both the regular distance ladder and the inverse distance ladder, it has been suggested that some SN Ia systematic may be responsible for the observed Hubble tension. However, this seems exceedingly unlikely: we show that SN Ia systematic effects investigated by Pantheon+ \citep{Brout22} and DES-SN5YR \citep{DES5YR, Popovic26} change $H_0$ by less than $0.1$ km\,s$^{-1}$ Mpc$^{-1}$, highlighting the precision of distance measurements with SNe Ia. On the contrary, SN Ia provide quite precise measurements of distance scales. \citet{Camarena23} suggests that there may not be an $H_0$ tension so much as an $M_0$ tension between low- and high-redshift absolute calibration sources, though this hypothesis is ruled out by the strong disagreement of the data with a magnitude step as in Equation \ref{eq:step}.

The $d\mu/dz$ that would be needed in SN Ia distance measurements to deviate from the CMB-set distance scale is on the order of $0.2$ mag. Such a change would be apparent in other distance measurements with SNe Ia, shifting the observed $\Omega_M$ measurements with SNe Ia into strong tension ($\sim10\sigma$) with other cosmological probes. It may be tempting to identify such a redshift evolution to the claimed \lq age bias\rq\ of SN Ia progenitors as outlined in \cite{Son25}. However, the evolution in SN Ia properties that would be needed to create this level of redshift drift is not detected in SN Ia samples \citep[see][for a summary]{Wiseman26}, and the SN Ia sample analysed in this paper has the magnitude evolution allowed by current data included as one of its systematic uncertainties -- $\gamma(z)$ in Fig.~\ref{fig:systematics}. 

Strongly-lensed SNe Ia \citep{Kelly23,Arendse24,Pierel25} provide a valuable, independent constraint, and gravitational waves \citep{LIGO17} show a promising avenue forwards for resolving the Hubble Tension as well. Happily, there exists many literature review papers on the topic: for TRGB measurements, see \cite{LiBeaton24}; for a more general overview, see \cite{Shah21}. \cite{Dhawan20, Dhawan25} provide a theoretical exploration of exotic dark energy theories and the distance-duality relationship, respectively. 

In summary, because of the relative consistency of distance measurements from current SN Ia datasets with the standard cosmological model, and the absolute distance scale set by the CMB, it is simple to make the prediction that any future measurements of $H_0$ with the inverse distance ladder calibrated by current CMB data will result in $H_0 = 66.5 \pm 0.5$ km\,s$^{-1}$ Mpc$^{-1}$.

\section*{Acknowledgments}

We would like to thank Robert Nichol, Edward Macaulay, Tamara Davis, Adam Riess, and Daniel Scolnic for helpful conversations. 

This project has received funding from the European Union’s Horizon Europe research and innovation programme under the Marie Skłodowska-Curie grant agreement No 101205780. MS acknowledges support from the Science and Technology Facilities Council (STFC) grant ST/Y001850/1.

\section{Data Availability}

The DES-Dovekie distances are available at \url{https://github.com/des-science/DES-SN5YR}, and the BAO distances are published in \cite{DESI25}.


\appendix

\section{Cosmographic Expansion In A Curved Universe}\label{sec:Appendix}

Following \cite{Visser04}, we lay out the cosmographic expansion of $\mathcal{D}_L(z)$ without the assumption of spatial flatness; there is no change to the $\mathcal{H}(z)$ term. The expansion is similar, with the relevant spatial $k$ term added to the $\mathcal{D}_2$ and $\mathcal{D}_3$ terms:
\begin{equation}
    D_L(z) = \frac{c}{H_0}\times(z + \mathcal{D}_1z^2 + \mathcal{D}_2z^3 + \mathcal{D}_3z^4 )
\end{equation}
where $\mathcal{D_i}$ are given by
\begin{equation}
    \mathcal{D}_1 = \frac{1}{2}(1-q_0),
\end{equation}
\begin{equation}
    \mathcal{D}_2 = -\frac{1}{6}(1-q_0 - 3q_0^2+j_0 + \frac{kc^2}{H_0^2a_0^2} ),
\end{equation}
\begin{align} 
\begin{split}
    \mathcal{D}_3 = \frac{1}{24}(2-2q_0-15q_0^2+5j_0+10q_0j_0 + s_0
    \\ + \frac{2kc^2(1+3q_0)}{H_0^2a_0^2}).
\end{split}
\end{align}
$a_0$ is the present-day value of the scale factor $a$. Given our interest in the consistency of the inverse distance ladder measurement of $H_0$, we marginalise the term $k/a_0^2$ rather than $k$.



\bibliographystyle{mnras}
\bibliography{research2, matt} 

\bsp	
\label{lastpage}
\end{document}